\documentclass[12pt]{article}
\usepackage{amssymb}
\usepackage{amsmath}
\usepackage[dvips]{graphicx}
\usepackage{pstricks}
\usepackage{pst-node}
\date{}
\title{Quantization of Games: Towards Quantum Artificial Intelligence}
\author{Katarzyna Miakisz \\ Institute of Mathematics,
University of Bia\l ystok,\\ Akademicka 2, Pl 15267 Bia\l ystok,
Poland\\e-mail: kmiakisz@math.uwb.edu.pl\\[1ex]
Edward W. Piotrowski\\ Institute of Mathematics,
University of Bia\l ystok,\\ Lipowa 41, Pl 15424 Bia\l ystok,
Poland\\ e-mail: ep@alpha.uwb.edu.pl \\[1ex]
 Jan S\l adkowski\\ Institute of Physics, University of Silesia, \\ Uniwersytecka
4, Pl 40007 Katowice, Poland \\ e-mail: sladk@us.edu.pl }
\begin{document}
\def\meter{\mbox{$\frown\hspace{-.9em}{\lower-.4ex\hbox{$_\nearrow$}}$}}
\def\circs{\mbox{$\hspace{-.3em}\bigcirc\hspace{-.69em}{\lower-.35ex\hbox{$_{S}$}}$}}
\def\circh{\mbox{$\hspace{-.16em}\bigcirc\hspace{-.76em}{\lower-.35ex\hbox{$_{H}$}}$}}
\maketitle
\begin{abstract}
\noindent  On grounds of the discussed material, we reason about
possible future development of quantum game theory and its impact
on information processing and  the emerging information society.
The idea of quantum artificial intelligence is explained.
\end{abstract}
\vspace{5mm}
\vspace{5mm}

\section{Introduction}

The construction of global information infrastructure caused one
of the main paradigm shifts in human history: information is
becoming a crucial if not the most important resource. The
scientific community has realized that information processing is a
physical phenomenon and that information theory is inseparable
from both applied and fundamental physics.  Investigation into the
physical aspects of information processing opened new perspectives
of computation, cryptography and  communication methods.  Very
often quantum approach  provides advantages over the classical
setting. Often the problem can be perceived as game and there are
examples  illustrating  methods of gaining an advantage over
``classical opponents'' by using quantum strategies
\cite{Mey}-\cite{inv}. Note that games against nature \cite{mil}
and quantum evolutionary games \cite{IT1} certainly include those
for which nature is quantum mechanical.  In these cases one can
hardly speak about rational agents or players. Nevertheless, as we
will show, sort of quantum artificial intelligence can be invoked
here. In this paper we would like to convince the reader that the
research on quantum game theory cannot be neglected because
present technological development suggest that sooner or later
someone would take full advantage of quantum theory and may use
quantum strategies to beat us at some realistic game. At present,
it is difficult to find out if human consciousness explores
quantum phenomena although it seems to be at least as mysterious
as the quantum world. Humans have been applying quantum
technologies more or less successfully since its discovery. Does
it mean that our intelligence is being transformed into quantum
artificial intelligence (cf. quantum anthropic principle as
formulated in \cite{qantpri})? Humans have already overcome
several natural limitations with help of artificial tools. Is
information processing waiting for its turn?

\section{Quantization of games}
\begin{figure}[h]
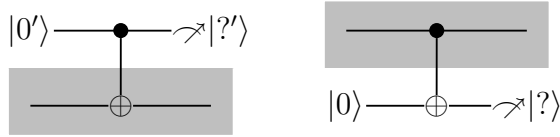

\begin{center}
\phantom{a}\vspace{5ex} \psset{linewidth=.7pt}
\rput(-3.3,1){\rnode{A}{$|0'\rangle$\hspace{1pt}}}
\cnode*(-2.1,1){.1}{B}
\rput(-0.9,1){\rnode{C}{\meter$|?'\rangle$}}
\pscustom[linecolor=white,fillstyle=solid,fillcolor=lightgray]{%
\psline(3.6,.5)(3.6,.5) \psline(3.6,1.4)(.6,1.4)
\psline(.6,1.4)(.6,.5) }
\rput(0.9,1){\rnode{D}{}} \cnode*(2.1,1){.1}{E}
\rput(3.3,1){\rnode{F}{}} \rput(-3.3,0){\rnode{G}{}}
\pscustom[linecolor=white,fillstyle=solid,fillcolor=lightgray]{%
\psline(-3.6,.5)(-3.6,.5) \psline(-3.6,-.4)(-.6,-.4)
\psline(-.6,-.4)(-.6,.5) }
\rput(-2.1,0){\rnode{H}{$\oplus$}} \rput(-0.9,0){\rnode{I}{}}
\rput(0.9,0){\rnode{J}{$|0\rangle$\hspace{1pt}}}
\rput(2.1,0){\rnode{K}{$\oplus$}}
\rput(3.3,0){\rnode{L}{\meter$|?\rangle$}}
\ncline[nodesep=0pt]{-}{A}{C} \ncline[nodesep=0pt]{-}{D}{F}
\ncline[nodesep=0pt]{-}{G}{H} \ncline[nodesep=0pt]{-}{H}{I}
\ncline[nodesep=0pt]{-}{J}{K} \ncline[nodesep=0pt]{-}{K}{L}
\ncline[nodesep=0pt]{-}{B}{H} \ncline[nodesep=0pt]{-}{E}{K}
\end{center}
\caption{The alliance as a mean of determining others' strategies.
The sign ``\meter\hspace{.06em}'' at the right ends of lines
representing qubits symbolizes measurement.} \label{sulokitek}
\end{figure}

Classical games usually cannot be quantized in a unique way
because they are only asymptotical ``shadows'' of a wide spectra of
quantum models. There are two obvious modifications of classical
simulation games.
\begin{description}
\item[1 -- prequantization:] Redefine the game so that it becomes
a reversal operation on qubits representing player's strategies.
This already allows for quantum coherence of
strategies\footnote{This may result from nonclassical initial
strategies or classically forbidden measurements of the state of
the game (end of the game).}. \item[2 -- quantization:] Reduce
\label{punktdwa} the number of qubits and allow arbitrary
unitary\footnote{At least one of the performed (allowed)
operations should not be equivalent to a classical one. Otherwise
we would get a game equivalent to some variant of the prequantized
classical game.} transformation so that the basic features of the
classical game are preserved. At this stage ancillary qubits can
be introduced so that possibly all quantum subtleties can be
explored (e.g.~entanglement, measurements and the involved
reductions of states, nonlocal quantum gates etc.).
\end{description}
One of the most appealing features of quantum games is the
possibility that  strategies can influence each other and form
collective strategies. Elsewhere \cite{komp}, we have defined the
alliance as  the gate CNOT ($\mathcal{C}$) regardless of its
standard name {\em controlled-NOT}\/  because it can be used to
form collective strategies as follows. Most of two-qubit quantum
gates are universal in the sense that any other gate can be
composed of a universal one \cite{DBE}-\cite{Lom}. Therefore it is
sufficient to describe a collective tactic of $N$ players as a
sequence of various operations $\mathcal{U}_{z,\alpha}$ belonging
to $SU(2)$ performed on one-dimensional subspaces of players'
strategies and, possibly, alliances $\mathcal{C}$ among them (any
element of $SU(2^N)$ can be given such a form \cite{BBC}).
Alliances are, up to equivalence, the only ways of forming
collective games. An alliance has the explicit form
$CNOT:=|0\rangle\langle0|\otimes I+ |1\rangle\langle1|\otimes
NOT$, where the tactic $NOT$ is represented in the qubit basis
$(|0\rangle,|1\rangle)$ by the matrix $\begin{pmatrix}0 &
\text{i}\\\text{i}&0\end{pmatrix}\negthinspace\in\negthinspace
SU(2)$. An alliance allows the player to determine the  state of
another player by entering into an alliance and measuring her
resulting strategy. This process is shortly described as
\begin{equation*}
\mathcal{C}\,|0'\rangle|m'\rangle=|m'\rangle|m'\rangle,\,\,\,\,
\mathcal{C}\,|m\rangle|0\rangle=|m\rangle|m\rangle,
\end{equation*}
where $m\negthinspace=\negthinspace0,\text{1}$. The corresponding
diagrams are shown in Fig\mbox{.} \ref{sulokitek}. The left
diagram presents measurement of the observable $\mathcal{X}'$ and
the right one  measurement of $\mathcal{X}$.
Any measurement would demolish possible entanglement of
strategies.  Therefore entangled quantum strategies can exist only
if the players in question  are ignorant of the details of their
strategies. To illustrate the problem we analyse three simple
games involving alliances. They can be used as partial solutions
in more complicated situations.
\section{Quantum solution to the Newcomb's paradox}
\begin{figure}[h]
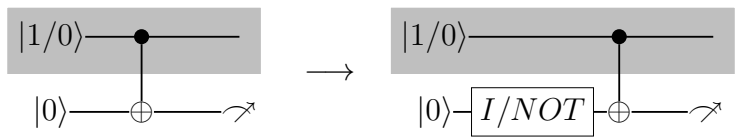

\phantom{.}\vspace{5ex}
\begin{center}
\pscustom[linecolor=white,fillstyle=solid,fillcolor=lightgray]{%
\psline(-4.9,.5)(-4.9,1.4) \psline(-4.9,1.4)(-1.5,1.4)
\psline(-1.5,.5)(-1.5,.5) }
\pscustom[linecolor=white,fillstyle=solid,fillcolor=lightgray]{%
\psline(0.2,.5)(0.2,.5) \psline(0.2,1.4)(4.8,1.4)
\psline(4.8,1.4)(4.8,.5) } \rput(-4.3,1){\rnode{A}{$|1/0\rangle$}}
\cnode*(-3.1,1){.1}{B} \rput(-1.8,1){\rnode{C}{}}
\rput(-4.3,0){\rnode{D}{$|0\rangle$}}
\rput(-3.1,0){\rnode{E}{$\oplus$}}
\rput(-1.8,0){\rnode{F}{\meter}}
\rput(-.6,.5){\rnode{G}{$\longrightarrow$}}
\rput(.8,1){\rnode{H}{$|1/0\rangle$}} \cnode*(3.25,1){.1}{I}
\rput(4.5,1){\rnode{J}{}} \rput(.8,0){\rnode{K}{$|0\rangle$}}
\rput(2.1,0){\rnode{L}{\psframebox[linewidth=.3pt]{$I/NOT$}}}
\rput(3.25,0){\rnode{M}{$\oplus$}} \rput(4.4,0){\rnode{O}{\meter}}
\psset{linewidth=.7pt} \ncline[nodesep=0pt]{-}{A}{C}
\ncline[nodesep=0pt]{-}{D}{E} \ncline[nodesep=0pt]{-}{E}{F}
\ncline[nodesep=0pt]{-}{B}{E}
\ncline[nodesep=0pt]{-}{H}{J} \ncline[nodesep=0pt]{-}{K}{L}
\ncline[nodesep=0pt]{-}{L}{M} \ncline[nodesep=0pt]{-}{M}{O}
\ncline[nodesep=0pt]{-}{I}{M}
\end{center}
\caption{Neutralization  of a quantum measuring system by a switch
$I/NOT$ applied ($I/NOT\rightarrow NOT$), when
$|1/0\rangle\negthinspace=\negthinspace|1\rangle$ (see the text).}
\label{grasweta}
\end{figure}

Let us consider the simple quantum circuit presented in
Fig\mbox{.} \ref{grasweta}. Any circuit is more or less vulnerable
to random errors. The gate $I/NOT$ is defined as a randomly chosen
gate from the set $\{I,NOT\}$) and is used to  switching-off the
circuit in a random way. It can be generalized to have some
additional control qubits. In a game-theoretical context such
circuits can be used to neutralization of disturbances caused by
measuring strategies, cf \cite{NC}.
For example, it can be applied to solve the famous Newcomb's free
will paradox \cite{PSN}. The problem, originally formulated by
William Newcomb in the 1960, was described by Martin Gardner in
the following way.  An alien Omega being a representative of alien
civilization (player 2) offers a human (player 1) a choice between
two boxes. The player 1 can take the content of both boxes or only
the content of the second one. The first one is transparent and
contains \$1000. Omega declares to have put into the second box
that is opaque \$1000000 (strategy $|\mathsf{1}\rangle_2$) but
only if Omega foresaw that the player 1 decided to take only the
content of that box ($|\mathsf{1}\rangle_1$). A male player 1
thinks: {\it If Omega knows what I am going to do then I have the
choice between \$1000 and \$1000000. Therefore I take the
\$1000000 }(strategy $|\mathsf{1}\rangle_1$). A female player 1
thinks: {\it Its obvious that I want to take the only the content
of the second box therefore Omega foresaw it and put the \$1000000
into the box. So the one million dollar {\bf is} in the second
box. Why should I not take more -- I take the content of both
boxes} (strategy $|\mathsf{0}\rangle_1$). The question is whose
strategy, male's or female's, is better? In he measuring system
presented in Fig\mbox{.} \ref{grasweta} the initial value
$|0\rangle$ of the lower qubit corresponds to the male strategy
and the values $|1\rangle$ and $|0\rangle$ of the upper qubit
correspond to male and female tactics, respectively. The outcome
$|0\rangle$ of a measurement performed on the lower qubit
indicates the opening of both boxes with contents prepared by
Omega before the alliance $CNOT$ was formed. If Omega  installed
in the circuit a breaker of the form $I/NOT$ (before or after the
alliance $CNOT$ he would use it when (and only then) the human
adopted the female tactics. But this would mean that Omega is
cheating (the breaker is installed after the alliance) or is able
to foretell the future (the breaker is installed before the
alliance). In the quantum setting the situation is different.
\begin{figure}[h]
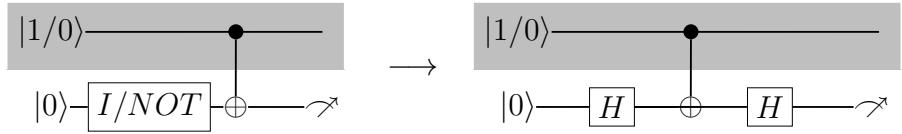

\phantom{.}\vspace{5ex}
\begin{center}
\pscustom[linecolor=white,fillstyle=solid,fillcolor=lightgray]{%
\psline(-6,.5)(-6,1.4) \psline(-6,1.4)(-1.5,1.4)
\psline(-1.5,.5)(-1.5,.5) }
\pscustom[linecolor=white,fillstyle=solid,fillcolor=lightgray]{%
\psline(.2,.5)(.2,.5) \psline(.2,1.4)(5.9,1.4)
\psline(5.9,1.4)(5.9,.5) }
\rput(-5.4,1){\rnode{A}{$|1/0\rangle$}} \cnode*(-2.95,1){.1}{B}
\rput(-1.8,1){\rnode{C}{}} \rput(-5.4,0){\rnode{D}{$|0\rangle$}}
\rput(-4.1,0){\rnode{E}{\psframebox[linewidth=.3pt]{$I/NOT$}}}
\rput(-2.95,0){\rnode{F}{$\oplus$}}
\rput(-1.8,0){\rnode{G}{\meter}}
\rput(-.6,.5){\rnode{H}{$\longrightarrow$}}
\rput(.8,1){\rnode{I}{$|1/0\rangle$}} \cnode*(3.1,1){.1}{J}
\rput(5.6,1){\rnode{K}{}} \rput(.8,0){\rnode{L}{$|0\rangle$}}
\rput(2.05,0){\rnode{M}{\psframebox[linewidth=.3pt]{$H$}}}
\rput(3.1,0){\rnode{N}{$\oplus$}}
\rput(4.15,0){\rnode{O}{\psframebox[linewidth=.3pt]{$H$}}}
\rput(5.5,0){\rnode{P}{\meter}}
\psset{linewidth=.7pt} \ncline[nodesep=0pt]{-}{A}{C}
\ncline[nodesep=0pt]{-}{D}{E} \ncline[nodesep=0pt]{-}{E}{F}
\ncline[nodesep=0pt]{-}{F}{G} \ncline[nodesep=0pt]{-}{B}{F}
\ncline[nodesep=0pt]{-}{I}{K} \ncline[nodesep=0pt]{-}{L}{M}
\ncline[nodesep=0pt]{-}{M}{N} \ncline[nodesep=0pt]{-}{M}{O}
\ncline[nodesep=0pt]{-}{O}{P} \ncline[nodesep=0pt]{-}{J}{N}
\end{center}
\caption{Solution to the Newcomb's paradox: quantum device that
neutralizes measurement. In the quantization process the gate
$I/NOT$ is replaced by a qutrojan (see the text) that acts
independently of the value of the qubit $|1/0\rangle$ and is
composed of  two Hadamard gates $H$.} \label{graswetcc}
\end{figure}
The quantization of the problem is presented in Fig.~\ref{graswetcc}.
It consists in replacing of the circuit-breaker
$I/NOT$ by a pair of Hadamard gates $H:=\frac{\text{i}}{\sqrt{2}}
\begin{pmatrix}1&\phantom{-}1\\
1&-1\end{pmatrix}\negthinspace\in\negthinspace SU(2)$. Due to
their jamming effect on the human's tactics, we can call them a
quantum Trojan horse (qutrojan)\footnote{Problems connected with
the definition of trojan are discussed in \cite{TAC}.}. We can hardy use
the term trojan with respect to the circuit-breaker $I/NOT$
because of its paradoxical correlation with human tactics. Note that
$H\cdot NOT \cdot H= \begin{pmatrix}-\text{i}&0\\
\phantom{-}0&\text{i}\end{pmatrix}$, hence any attempt at
measuring squared absolute values of coordinates of the human
strategy qubit will not detect any effectiveness of the female
tactics.

\section{Quantum Metropolis algorithm}
\begin{figure}[h]
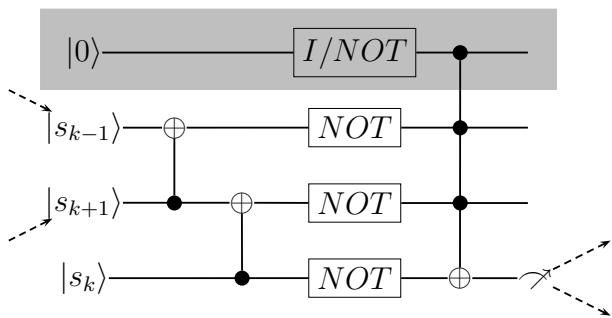

\phantom{a}\vspace{16ex}
\begin{center}
\pscustom[linecolor=white,fillstyle=solid,fillcolor=lightgray]{%
\psline(-3.6,2.5)(-3.6,3.6) \psline(-3.6,3.6)(3.3,3.6)
\psline(3.3,3.6)(3.3,2.5) } \rput(-3,3){\rnode{A}{$|0\rangle$}}
\rput(.6,3){\rnode{B}{\psframebox[linewidth=.3pt]{$I/NOT$}}}
\cnode*(2,3){.1}{C} \rput(2.9,3){\rnode{D}{}}
\rput(-3,2){\rnode{E}{$|s_{k-1}\rangle$}}
\rput(-4.,2.5){\rnode{V}{}} \rput(-1.8,2){\rnode{F}{$\oplus$}}
\rput(.6,2){\rnode{G}{\psframebox[linewidth=.3pt]{$NOT$}}}
\cnode*(2,2){.1}{H} \rput(2.9,2){\rnode{I}{}}
\rput(-3,1){\rnode{J}{$|s_{k+1}\rangle$}}
\rput(-4.,.5){\rnode{W}{}} \cnode*(-1.8,1){.1}{K}
\rput(-.9,1){\rnode{L}{$\oplus$}}
\rput(.6,1){\rnode{M}{\psframebox[linewidth=.3pt]{$NOT$}}}
\cnode*(2,1){.1}{N} \rput(2.9,1){\rnode{O}{}}
\rput(-3,0){\rnode{P}{$|s_{k}\rangle$}} \cnode*(-.9,0){.1}{R}
\rput(.6,0){\rnode{S}{\psframebox[linewidth=.3pt]{$NOT$}}}
\rput(2,0){\rnode{T}{$\oplus$}} \rput(3,0){\rnode{U}{\meter}}
\rput(4.,.5){\rnode{X}{}}%
\rput(4.,-.5){\rnode{Y}{}}%
\psset{linewidth=.7pt} \ncline[nodesep=0pt]{-}{A}{B}
\ncline[nodesep=0pt]{-}{B}{D} \ncline[nodesep=0pt]{-}{E}{F}
\ncline[nodesep=0pt]{-}{F}{G} \ncline[nodesep=0pt]{-}{G}{I}
\ncline[nodesep=0pt]{-}{J}{L} \ncline[nodesep=0pt]{-}{L}{M}
\ncline[nodesep=0pt]{-}{M}{O} \ncline[nodesep=0pt]{-}{P}{S}
\ncline[nodesep=0pt]{-}{S}{T} \ncline[nodesep=0pt]{-}{T}{U}
\ncline[nodesep=0pt]{-}{F}{K} \ncline[nodesep=0pt]{-}{L}{R}
\ncline[nodesep=0pt]{-}{C}{T}
\ncline[nodesep=0pt,linestyle=dashed,dash=3pt 2pt]{->}{V}{E}
\ncline[nodesep=0pt,linestyle=dashed,dash=3pt 2pt]{->}{W}{J}
\ncline[nodesep=0pt,linestyle=dashed,dash=3pt 2pt]{->}{U}{X}
\ncline[nodesep=0pt,linestyle=dashed,dash=3pt 2pt]{->}{U}{Y}
\end{center}
\caption{Tactics of the $k$-th cell of a network simulating the
Metropolis algorithm (Ising chain).} \label{grymeertpo}
\end{figure}

An obvious generalization of the CNOT gate consists in adding more
control bits. Let us consider a cellular automaton that is able to
implement the popular Metropolis algorithm \cite{metropolis}. Such
an automaton can be constructed by forming a network of identical
sub-automata (that is implementing the same tactics) joined by
classical communication channels. Were the communication channels
quantum (that is admitting nonlocal alliances), the system formed
by automata implementing arbitrary one qubit tactics would be a
fully-fledged quantum computer of distributed architecture
\cite{lomonaco2}. In order to eliminate the possible feedback
catastrophe  in single simulation step only part of the cells
should be activated \cite{vichniac}. Let us restrict ourselves to
(local) quantization  procedure such that  only the sub-automata
are acted on. To simulate the $1$D  Ising model\footnote{An
extended description of such simulation of the Ising model can be
found in the paper \cite{cole} were a more complicated automaton
is used to this end.} the network has the cyclic group $Z^N$
structure and the automaton can be built in the form presented in
Fig.~\ref{grymeertpo}.

The sub-automaton (cell) is built in such a way that the
activation does not change its strategy $|s_k\rangle$
($I/NOT\negthinspace\rightarrow\negthinspace NOT$ what happens
with the probability $p$) only if the strategies of the
neighboring cells $|s_{k-1}\rangle$ and $|s_{k+1}\rangle$ have the
same strategies. The dashed lines represent one bit information
flows between neighboring cells. A simple quantization of this
system that does not influence results of the simulation consists
in replacing the switch $I/NOT$ with such a one-qubit tactics $U$
that $|\langle 1|U|0\rangle|^2\negthinspace=\negthinspace p$. The
quantization results in elimination of a time-consuming
pseudo-random numbers generator that is necessary for correct
performance of the switch $I/NOT$. The sub-automaton can be
rebuilt so that the network will simulate more dimensional Ising
model \cite{cole}. Going farther in this direction, by choosing
for the measurement basis (in a preselected or random way) for the
strategy $|s_k\rangle$ various conjugated bases that are
equivalent to additional one-qubit tactics\footnote{To perform
this measurement, one simply unitarily transforms from the basis
we wish to perform the measurement into the computational basis,
then measure.} \cite{wiesner} we will be able, for example, to
simulate the evolution of cliques that might form in quantum
market games \cite{PS1, komp}. Note that this sort of a quantum
version of the Metropolis algorithm can be effectively implemented
on a classical computer.
\section{The Elitzur--Vaidman circuit--breaker}
\begin{figure}[h]
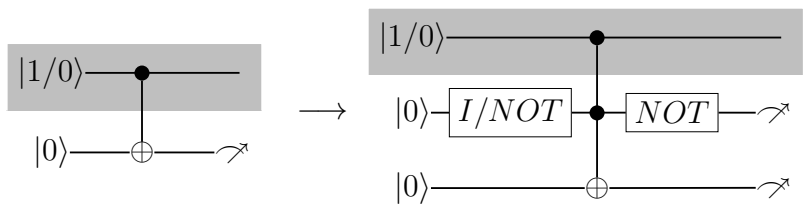

\phantom{a}\vspace{11ex}
\begin{center}
\pscustom[linecolor=white,fillstyle=solid,fillcolor=lightgray]{%
\psline(-5.4,1.03)(-5.4,1.93) \psline(-5.4,1.93)(-2,1.93)
\psline(-2,1.93)(-2,1.03) }
\pscustom[linecolor=white,fillstyle=solid,fillcolor=lightgray]{%
\psline(-.6,1.5)(-.6,1.5) \psline(-.6,2.4)(5.2,2.4)
\psline(5.2,2.4)(5.2,1.5) }
\rput(-4.8,1.53){\rnode{A}{$|1/0\rangle$}}
\cnode*(-3.6,1.53){.1}{B} \rput(-2.3,1.53){\rnode{C}{}}
\rput(-4.8,.47){\rnode{D}{$|0\rangle$}}
\rput(-3.6,.47){\rnode{E}{$\oplus$}}
\rput(-2.4,.47){\rnode{F}{\meter}}
\rput(-1.2,1){\rnode{G}{$\longrightarrow$}}
\rput(0,2){\rnode{H}{$|1/0\rangle$}} \cnode*(2.45,2){.1}{I}
\rput(4.9,2){\rnode{J}{}} \rput(0,1){\rnode{K}{$|0\rangle$}}
\rput(1.3,1){\rnode{L}{\psframebox[linewidth=.3pt]{$I/NOT$}}}
\cnode*(2.45,1){.1}{M}
\rput(3.45,1){\rnode{N}{\psframebox[linewidth=.3pt]{$NOT$}}}
\rput(4.8,1){\rnode{O}{\meter}} \rput(0,0){\rnode{P}{$|0\rangle$}}
\rput(2.45,0){\rnode{Q}{$\oplus$}} \rput(4.8,0){\rnode{R}{\meter}}
\psset{linewidth=.7pt} \ncline[nodesep=0pt]{-}{A}{C}
\ncline[nodesep=0pt]{-}{D}{E} \ncline[nodesep=0pt]{-}{E}{F}
\ncline[nodesep=0pt]{-}{B}{E}
\ncline[nodesep=0pt]{-}{H}{J} \ncline[nodesep=0pt]{-}{K}{L}
\ncline[nodesep=0pt]{-}{L}{N} \ncline[nodesep=0pt]{-}{N}{O}
\ncline[nodesep=0pt]{-}{P}{Q} \ncline[nodesep=0pt]{-}{Q}{R}
\ncline[nodesep=0pt]{-}{I}{Q}
\end{center}
\caption{Modification of the system by adding a switching-off
strategy.} \label{grytuty}
\end{figure}

Let us now consider a modification of the method of jamming of the
strategy measuring game in which the circuit-breaker gate  $I/NOT$
is implemented as a part  in a separate switching-off strategy,
cf.~Fig\mbox{.} \ref{grytuty}. To this end the alliance  $CNOT$
was replaced by the Toffoli gate (Controlled-Controlled-NOT).
Contrary to former case we are now interested in effective
accomplishment of the measurement. Therefore we assume that there
are no correlations between the state of the gate $I/NOT$ and the
strategy $|1/0\rangle$. The role of the gate $NOT$ that comes
before the measurement of the central qubit is to guarantee that
the measurement of the state  $|1\rangle$ stands for the
switching-off of the subsystem consisting of the two bottom
qubits.
To quantize this game we will follow  Elitzur and Vaidman
\cite{Penrose} who explored Mauritius Renninger's idea of the {\it
negative measurement} \cite{renninger}, see Fig\mbox{.}
\ref{gryt}. The method consists in gradual unblocking of the
switching-off strategy ($n$ steps of $\sqrt[n]{NOT}$) and at each
step, if only the change of the third qubit (measuring the first
qubit) is observed, the whole measurement is given up. So the game
is stopped by the ``exploding bomb''\footnote{Note that exploding
bombs can actually be priceless for implementations quantum
algorithms, cf\mbox{.} \cite{rudolph}.} what happens if at some
step the value of the auxiliary strategy measured after the
alliance $CNOT$ is measured to be $|1\rangle$, see Fig\mbox{.}
\ref{gryt}.
\newcommand{\bang}{%
\psset{linewidth=.3pt}
\begin{pspicture}[0](0,0)(1.2,.8)
\pspolygon[shadow=true,shadowsize=1.1pt,shadowangle=-120]%
(.00,.40)(.28,.44)(.16,.60)(.40,.52)(.44,.72)(.60,.56)(.64,.80)(.76,.64)
(.88,.72)(.88,.56)(1.12,.64)(1.0,.52)(1.2,.44)(.96,.36)(1.16,.2)(.84,.28)
(.8,.16)(.64,.24)(.48,.00)(.44,.20)(.24,.12)(.32,.24)(.08,.20)(.24,.32)
\rput(.61,.4){\rnode{V}{\sc\tiny bang!!}}
\end{pspicture}}
\begin{figure}[h]
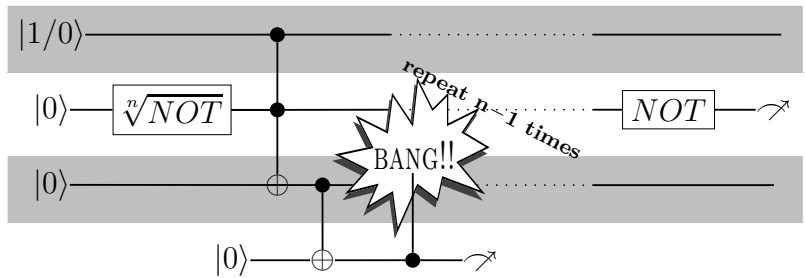

\begin{center}
\phantom{a}\vspace{16ex}
\pscustom[linecolor=white,fillstyle=solid,fillcolor=lightgray]{%
\psline(-5.4,2.5)(-5.4,3.4) \psline(-5.4,3.4)(5.2,3.4)
\psline(5.2,3.4)(5.2,2.5) }
\pscustom[linecolor=white,fillstyle=solid,fillcolor=lightgray]{%
\psline(-5.4,0.5)(-5.4,1.4) \psline(-5.4,1.4)(5.2,1.4)
\psline(5.2,1.4)(5.2,0.5) }
\rput(-4.8,3){\rnode{A}{$|1/0\rangle$}}\rput(-3.6,3){\rnode{B}{}}
\cnode*(-1.8,3){.1}{C}\rput(-.8,3){\rnode{D}{}}
\rput(-.3,3){\rnode{F}{}}\rput(2.4,3){\rnode{G}{}}
\rput(3.6,3){\rnode{H}{}}\rput(4.9,3){\rnode{I}{}}
\rput(-4.8,2){\rnode{A1}{$|0\rangle$}}
\rput(-3.2,2){\rnode{B1}{\psframebox[linewidth=.3pt]{\phantom{.}\hspace{-.4em}$\sqrt[n]{NOT}$}}}
\cnode*(-1.8,2){.1}{C1}\rput(-.8,2){\rnode{D1}{}}
\rput(-.3,2){\rnode{F1}{}}\rput(2.4,2){\rnode{G1}{}}
\rput(3.4,2){\rnode{H1}{\psframebox[linewidth=.3pt]{$NOT$}}}
\rput(4.8,2){\rnode{I1}{\meter}}
\rput(-4.8,1){\rnode{A2}{$|0\rangle$}}\rput(-3.6,1){\rnode{B2}{}}
\rput(-1.8,1){\rnode{C2}{$\oplus$}}\cnode*(-1.2,1){.1}{D2}
\rput(-.3,1){\rnode{F2}{}}\rput(2.4,1){\rnode{G2}{}}
\rput(3.6,1){\rnode{H2}{}}\rput(4.8,1){\rnode{I2}{}}
\rput(-2.4,0){\rnode{X}{$|0\rangle$}}\rput(-1.2,0){\rnode{Y}{$\oplus$}}
\cnode*(.0,0){.1}{Z} \rput(.0,1.16){\rnode{T}{}}
\rput(.9,0){\rnode{V}{\meter}}
\psset{linewidth=.7pt} \ncline[nodesep=0pt]{-}{C}{C2}
\ncline[nodesep=0pt]{-}{D2}{Y}
\ncline[nodesep=0pt]{-}{A}{F}\ncline[nodesep=0pt,linestyle=dotted]{-}{F}{G}
\ncline[nodesep=0pt]{-}{G}{I}
\ncline[nodesep=0pt]{-}{A1}{B1}\ncline[nodesep=0pt]{-}{B1}{C1}
\ncline[nodesep=0pt]{-}{C1}{F1}\ncline[nodesep=0pt]{-}{B1}{C1}
\ncline[nodesep=0pt,linestyle=dotted]{-}{F1}{G1}\ncline[nodesep=0pt]{-}{G1}{H1}\ncline[nodesep=0pt]{-}{H1}{I1}
\ncline[nodesep=0pt]{-}{A2}{F2}\ncline[nodesep=0pt,linestyle=dotted]{-}{F2}{G2}\ncline[nodesep=0pt]{-}{G2}{I2}
\ncline[nodesep=0pt]{-}{X}{Y}\ncline[nodesep=0pt]{-}{Y}{Z}
\ncline[nodesep=0pt]{-}{Z}{V} \psset{linewidth=.0pt}
\pcline[linestyle=none](2.4,1.3)(-.3,2.7) \lput{:D}{\scriptsize\bf
repeat $\mathbf{n\negthinspace-\negthinspace1}$ times}
\rput[fillstyle=solid](0,1.4){\rnode{TT}{\scaleboxto(2,2){\bang}}}
\psset{linewidth=.7pt} \ncline[nodesep=0pt]{-}{Z}{T}
\end{center}\vspace{2ex}
\caption{The Elitzur--Vaidman tactics of gradual unblocking of the
switching-off strategy.} \label{gryt}
\end{figure}
The tactics $\sqrt[n]{NOT}$ of gradual unblocking is represented
by the operator:
\begin{equation*}
\sqrt[n]{NOT}:=I\cos\tfrac{\pi}{2n} +NOT\sin\tfrac{\pi}{2n}
 =\text{e}^{NOT\tfrac{\pi}{2n}}\negthinspace\in\negthinspace SU(2)\,.
\end{equation*}
The probability of continuation of the game after one step is
equal to $$\bigl|\langle0|\sqrt[n]{NOT}|0\rangle\bigr|^2=
\cos^2(\tfrac{\pi}{2n})$$ and all steps are successfully
accomplished with probability
$\cos^{2n}(\tfrac{\pi}{2n})=1-\tfrac{\pi^2}{4n}+\tfrac{\pi^4}{32n^2}+O(n^{-3})$.
Therefore in the limit
$n\negthinspace\rightarrow\negthinspace\infty$  the probability of
stopping the game tends to zero\footnote{The limit can be found by
application of the de L'Hospital rule to
$\ln\cos^{2n}\tfrac{\pi}{2n}$.}. The inspection of the value of
the first qubit with help of the third qubit gets a transcendental
dimension because if
$|1/0\rangle\negthinspace=\negthinspace|1\rangle$ the measuring
system is switched-off and if
$|1/0\rangle\negthinspace=\negthinspace|0\rangle$ the
switching-off strategy cannot be unblocked. The bomb plays the key
role in the game because it freezes the second qubit in the state
$|0\rangle$ --- this is the famous quantum Zeno effect
\cite{facchi}. But the information about the state of the first
qubit ($|0\rangle$ or $|1\rangle$) can only be acquired via the
effectiveness of the unblocking of the second qubit. The presented
implementation and analysis of the Elitzur-Vaidman circuit-breaker
paves the way for a completely new class of technologies that
might be shocking for those unacquainted with quantum effects. For
example, if the first qubit represents a result of quantum
computation then such a breaker allows to access in that part of
the Deutsch Multiversum \cite{deutsch2} where this computer is
turned off \cite{mitchison}. If the first qubit of the circuit
presented in Fig.~\ref{gryt} is fixed in the state $|1\rangle$
then this machinery can be to no demolition measurement that, for
example,  is able to select bombs with damaged fuse. The
respective measuring system is presented in Fig.~\ref{grytane}
(the shaded-in qubits in Fig.~\ref{gryt} are absent because they
are redundant). The breaker $Controlled(I/NOT)$ that replaces the
alliance $CNOT$ is in the state $I/NOT\negthinspace=\negthinspace
I$ if the bomb fuse is damaged and in the state
$I/NOT\negthinspace=\negthinspace NOT$ if the fuse is working.
\begin{figure}[h]
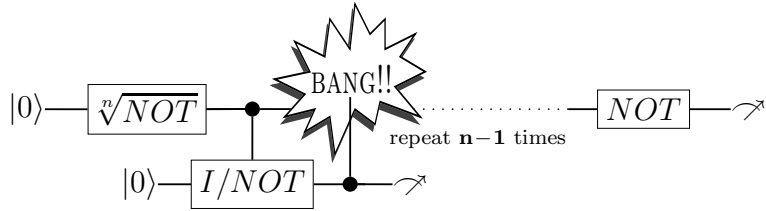

\begin{center}
\phantom{a}\vspace{16ex}
\rput(-4.8,2){\rnode{A1}{$|0\rangle$}}
\rput(-3.2,2){\rnode{B1}{\psframebox[linewidth=.3pt]{\phantom{.}\hspace{-.4em}$\sqrt[n]{NOT}$}}}
\cnode*(-1.8,2){.1}{C1}\rput(-.8,2){\rnode{D1}{}}
\rput(-.3,2){\rnode{F1}{}}\rput(2.4,2){\rnode{G1}{}}
\rput(3.4,2){\rnode{H1}{\psframebox[linewidth=.3pt]{$NOT$}}}
\rput(4.8,2){\rnode{I1}{\meter}}
\rput(-3.3,1){\rnode{A2}{$|0\rangle$}}
\rput(-1.8,1){\rnode{C2}{\psframebox[linewidth=.3pt]{$I/NOT$}}}
\rput(-.3,1){\rnode{F2}{}}\rput(2.4,1){\rnode{G2}{}}
\rput(3.6,1){\rnode{H2}{}}\rput(4.8,1){\rnode{I2}{}}
\cnode*(-.5,1){.1}{Z} \rput(-.5,2.16){\rnode{T}{}}
\rput(.3,1){\rnode{V}{\meter}}
\psset{linewidth=.7pt} \ncline[nodesep=0pt]{-}{C}{C2}
\ncline[nodesep=0pt]{-}{D2}{Y}
\ncline[nodesep=0pt]{-}{A1}{B1}\ncline[nodesep=0pt]{-}{B1}{C1}
\ncline[nodesep=0pt]{-}{C1}{F1}\ncline[nodesep=0pt]{-}{B1}{C1}
\ncline[nodesep=0pt]{-}{C1}{C2}
\ncline[nodesep=0pt,linestyle=dotted]{-}{F1}{G1}\ncline[nodesep=0pt]{-}{G1}{H1}\ncline[nodesep=0pt]{-}{H1}{I1}
\ncline[nodesep=0pt]{-}{A2}{C2}\ncline[nodesep=0pt]{-}{C2}{F2}
\ncline[nodesep=0pt]{-}{X}{Y}\ncline[nodesep=0pt]{-}{Y}{Z}
\ncline[nodesep=0pt]{-}{Z}{V} \psset{linewidth=.0pt}
\pcline[linestyle=none](2.4,1.6)(0,1.6) \lput{:D}{\scriptsize
repeat $\mathbf{n\negthinspace-\negthinspace1}$ times}
\rput[fillstyle=solid](-.5,2.4){\rnode{TT}{\scaleboxto(2,2){\bang}}}
\psset{linewidth=.7pt} \ncline[nodesep=0pt]{-}{Z}{T}\vspace{-7ex}
\end{center}
\caption{Safe Elitzur--Vaidman  bomb tester.} \label{grytane}
\end{figure} The result $|1\rangle$ of measurement of the first qubit tell us that the bomb is
in the working order. This is so because the working bomb always
reduces this qubit to  $|0\rangle$ after the transformation
$\sqrt[n]{NOT}$ (quantum Zeno effect). Of course such a bomb
tester (and the Elitzur--Vaidman circuit--breaker) can be
constructed on the basis of the quantum anti-Zeno effect
\cite{facchi2}. In this case the working but unexploded bomb
accelerates the evolution of the system instead of ``freezing'' it.
Such  alternative tester is presented in Fig\mbox{.}
\ref{grytanezer}, where the working bomb causes at any of the $n$
stages
\begin{figure}[h]
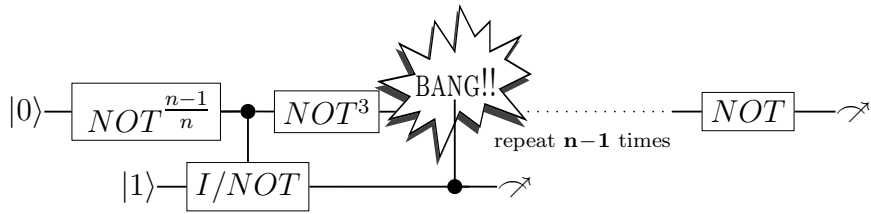

\begin{center}
\phantom{a}\vspace{16ex}
\rput(-5.5,2){\rnode{A1}{$|0\rangle$}}
\rput(-3.9,2){\rnode{B1}{\psframebox[linewidth=.3pt]{\phantom{.}\hspace{-.4em}
$NOT^{\tfrac{n-1}{n}}$}}} \cnode*(-2.55,2){.1}{C1}
\rput(-1.5,2){\rnode{Y1}{\psframebox[linewidth=.3pt]{$NOT^3$}}}
\rput(.4,2){\rnode{F1}{}}\rput(3.1,2){\rnode{G1}{}}
\rput(4.1,2){\rnode{H1}{\psframebox[linewidth=.3pt]{$NOT$}}}
\rput(5.5,2){\rnode{I1}{\meter}}
\rput(-4.,1){\rnode{A2}{$|1\rangle$}}
\rput(-2.55,1){\rnode{C2}{\psframebox[linewidth=.3pt]{$I/NOT$}}}
\rput(.4,1){\rnode{F2}{}}\rput(3.1,1){\rnode{G2}{}}
\rput(4.3,1){\rnode{H2}{}}\rput(5.5,1){\rnode{I2}{}}
\cnode*(.2,1){.1}{Z} \rput(.2,2.16){\rnode{T}{}}
\rput(1.,1){\rnode{V}{\meter}}
\psset{linewidth=.7pt} \ncline[nodesep=0pt]{-}{C}{C2}
\ncline[nodesep=0pt]{-}{D2}{Y}
\ncline[nodesep=0pt]{-}{A1}{B1}\ncline[nodesep=0pt]{-}{B1}{C1}
\ncline[nodesep=0pt]{-}{C1}{Y1} \ncline[nodesep=0pt]{-}{Y1}{F1}
\ncline[nodesep=0pt]{-}{B1}{C1} \ncline[nodesep=0pt]{-}{C1}{C2}
\ncline[nodesep=0pt,linestyle=dotted]{-}{F1}{G1}\ncline[nodesep=0pt]{-}{G1}{H1}\ncline[nodesep=0pt]{-}{H1}{I1}
\ncline[nodesep=0pt]{-}{A2}{C2}\ncline[nodesep=0pt]{-}{C2}{F2}
\ncline[nodesep=0pt]{-}{X}{Y}\ncline[nodesep=0pt]{-}{Y}{Z}
\ncline[nodesep=0pt]{-}{Z}{V} \psset{linewidth=.0pt}
\pcline[linestyle=none](3.1,1.6)(0.7,1.6) \lput{:D}{\scriptsize
repeat $\mathbf{n\negthinspace-\negthinspace1}$ times}
\rput[fillstyle=solid](.2,2.4){\rnode{TT}{\scaleboxto(2,2){\bang}}}
\psset{linewidth=.7pt} \ncline[nodesep=0pt]{-}{Z}{T}\vspace{-7ex}
\end{center}
\caption{Bomb tester constructed on the basis of the quantum
anti-Zeno effect.} \label{grytanezer}
\end{figure} the increase  of $\tfrac{\pi}{2n}$ in the phase $\varphi$ of the
cumulative tactics $\text{e}^{NOT\varphi}$. Let us define
$V(\beta):= NOT \cos\beta +( I \cos\alpha + H\cdot NOT \cdot H
\sin\alpha)\sin\beta$. It is easy to show that $V(\beta_2) \cdot
NOT^3 \cdot V(\beta_1)=V(\beta_1+\beta_2)$.  Therefore we can
replace the gate $NOT^{\tfrac{n-1}{n}}$ with any of the gates $$NOT
\cos\tfrac{\pi}{2n} +( I \cos\alpha + H\cdot NOT \cdot H
\sin\alpha)\sin\tfrac{\pi}{2n}\,,$$
 where
$\alpha\negthinspace\in\negthinspace[0,2\pi)$. But only for
$\alpha\negthinspace=\negthinspace0,\pi$ such gate belongs to the
class $\text{e}^{NOT\varphi}$  and  we can claim that the
transformation $NOT$ results from the acceleration or freezing of
the evolution of the system. For
$\alpha\negthinspace\neq\negthinspace0,\pi$ we observe sort of
para-Zeno effect because the measurement of the entangled with the
qubit in question qubit stops the free evolution corresponding to
a damaged bomb.
\begin{figure}[h]
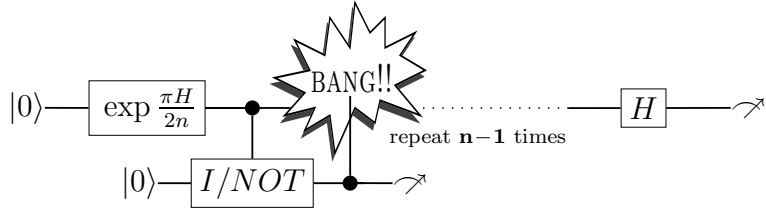

\begin{center}
\phantom{a}\vspace{14ex}
\rput(-4.8,2){\rnode{A1}{$|0\rangle$}}
\rput(-3.2,2){\rnode{B1}{\psframebox[linewidth=.3pt]{\phantom{.}\hspace{-.4em}
$\exp\tfrac{\pi H}{2n}$}}}
\cnode*(-1.8,2){.1}{C1}\rput(-.8,2){\rnode{D1}{}}
\rput(-.3,2){\rnode{F1}{}}\rput(2.4,2){\rnode{G1}{}}
\rput(3.4,2){\rnode{H1}{\psframebox[linewidth=.3pt]{$H$}}}
\rput(4.8,2){\rnode{I1}{\meter}}
\rput(-3.3,1){\rnode{A2}{$|0\rangle$}}
\rput(-1.8,1){\rnode{C2}{\psframebox[linewidth=.3pt]{$I/NOT$}}}
\rput(-.3,1){\rnode{F2}{}}\rput(2.4,1){\rnode{G2}{}}
\rput(3.6,1){\rnode{H2}{}}\rput(4.8,1){\rnode{I2}{}}
\cnode*(-.5,1){.1}{Z} \rput(-.5,2.16){\rnode{T}{}}
\rput(.3,1){\rnode{V}{\meter}}
\psset{linewidth=.7pt} \ncline[nodesep=0pt]{-}{C}{C2}
\ncline[nodesep=0pt]{-}{D2}{Y}
\ncline[nodesep=0pt]{-}{A1}{B1}\ncline[nodesep=0pt]{-}{B1}{C1}
\ncline[nodesep=0pt]{-}{C1}{F1}\ncline[nodesep=0pt]{-}{B1}{C1}
\ncline[nodesep=0pt]{-}{C1}{C2}
\ncline[nodesep=0pt,linestyle=dotted]{-}{F1}{G1}\ncline[nodesep=0pt]{-}{G1}{H1}\ncline[nodesep=0pt]{-}{H1}{I1}
\ncline[nodesep=0pt]{-}{A2}{C2}\ncline[nodesep=0pt]{-}{C2}{F2}
\ncline[nodesep=0pt]{-}{X}{Y}\ncline[nodesep=0pt]{-}{Y}{Z}
\ncline[nodesep=0pt]{-}{Z}{V} \psset{linewidth=.0pt}
\pcline[linestyle=none](2.4,1.6)(0,1.6) \lput{:D}{\scriptsize
repeat $\mathbf{n\negthinspace-\negthinspace1}$ times}
\rput[fillstyle=solid](-.5,2.4){\rnode{TT}{\scaleboxto(2,2){\bang}}}
\psset{linewidth=.7pt} \ncline[nodesep=0pt]{-}{Z}{T}\vspace{-7ex}
\end{center}
\caption{Supply-demand switch.} \label{osrede}
\end{figure} Consider a slight modification of the circuit presented in Fig.~\ref{osrede},
where now $\exp\tfrac{\pi H}{2n}= I\,\cos \tfrac{\pi }{2n}\ +
H\,\sin \tfrac{\pi}{2n}$. Again, we can avoid explosion  with a
high probability because
$(\,|\cos\tfrac{\pi}{2n}+\tfrac{\text{i}}{\sqrt{2}}\sin\tfrac{\pi}{2n}\,|^2\,)^n\,
>\, \cos^{2n}\negthinspace\tfrac{\pi}{2n}$. In this case the
information revealed by the breaker is more subtle because the
``bomb'' can only cause transition to a corresponding state in the
conjugated basis \cite{wiesner}. Nevertheless, the bomb being in
the working order cases  the strategy change. For example, in
quantum market games \cite{komp} models the supply strategy is
changed to the demand one. Such a mechanism can be used to
stabilize prices on a futuristic quantum markets if the market
crash is yet only a menace: an instability verifiable only by
provoking counterfactual crash.
\section{Identification of strategies}
\begin{figure}[h]
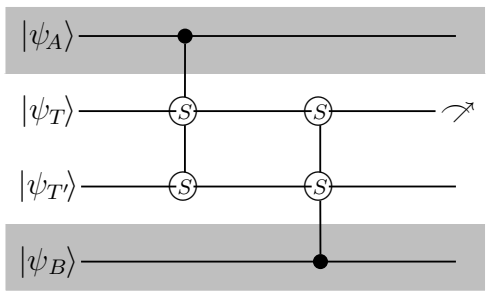

\begin{center}
\phantom{a}\vspace{15ex} \psset{linewidth=.7pt}
\pscustom[linecolor=white,fillstyle=solid,fillcolor=lightgray]{%
\psline(-3.3,2.5)(-3.3,2.5) \psline(-3.3,3.4)(3.1,3.4)
\psline(3.1,3.4)(3.1,2.5) }
\rput(-2.7,3){\rnode{A}{$|\psi_{\negthinspace
A}\rangle$\hspace{1pt}}} \cnode*(-.9,3){.1}{B}
\rput(2.7,3){\rnode{C}{}}
\rput(-2.7,2){\rnode{D}{$|\psi_{T}\hspace{-.1em}\rangle$\hspace{1pt}}}
\rput(-.9,2){\rnode{E}{\circs}} \rput(.9,2){\rnode{F}{\circs}}
\rput(2.7,2){\rnode{G}{\hspace{.15em}\meter}}
\rput(-2.7,1){\rnode{H}{$|\psi_{T'}\negthinspace\rangle$\hspace{1pt}}}
\rput(-.9,1){\rnode{I}{\circs}} \rput(.9,1){\rnode{J}{\circs}}
\rput(2.7,1){\rnode{K}{}}
\pscustom[linecolor=white,fillstyle=solid,fillcolor=lightgray]{%
\psline(-3.3,.5)(-3.3,.5) \psline(-3.3,-.4)(3.1,-.4)
\psline(3.1,-.4)(3.1,.5) }
\rput(-2.7,0){\rnode{L}{$|\psi_{\hspace{.15em}\negthinspace
B}\hspace{-.05em}\rangle$\hspace{1pt}}} \cnode*(.9,0){.1}{M}
\rput(2.7,0){\rnode{N}{}} \ncline[nodesep=0pt]{-}{A}{C}
\ncline[nodesep=0pt]{-}{D}{E} \ncline[nodesep=0pt]{-}{E}{F}
\ncline[nodesep=0pt]{-}{F}{G} \ncline[nodesep=0pt]{-}{H}{I}
\ncline[nodesep=0pt]{-}{I}{J} \ncline[nodesep=0pt]{-}{J}{K}
\ncline[nodesep=0pt]{-}{L}{N} \ncline[nodesep=0pt]{-}{B}{E}
\ncline[nodesep=0pt]{-}{E}{I} \ncline[nodesep=0pt]{-}{F}{J}
\ncline[nodesep=0pt]{-}{J}{M}
\end{center}
\caption{Identification game constructed from two
controlled--swap gates.} \label{goresyt1}
\end{figure}

Actually, any quantum computation is a potential quantum game if
we manage to reinterpret it in game--theoretical terms.
Identification of strategies is often a challenge in such process.
To illustrate the  method let us consider Wiesner's
counterfeit--proof banknote \cite{wiesner}. This is the first
quantum secrecy method (elimination of effective eavesdropping).
As a quantum game it consists in a finite series of sub--games
presented in Fig\mbox{.} \ref{goresyt1}. The arbiter Trent
produces a pair of random qubits $|\psi_{T}\hspace{-.1em}\rangle$
and $|\psi_{T'}\negthinspace\rangle$. The polarization of the
qubit (strategy) $|\psi_{T}\hspace{-.1em}\rangle$ is known to
Trent and is kept secret. The qubit
$|\psi_{T'}\negthinspace\rangle$ is ancillary.  Alice qubit
$|\psi_{\negthinspace A}\rangle$ describes her strategies
$|\text{I}\rangle$ and $|0\rangle$. The first move is performed by
Alice. Her strategy $|\text{I}\rangle$ consists in switching the
Trent's qubits $|\psi_{T}\hspace{-.1em}\rangle$ and
$|\psi_{T'}\negthinspace\rangle$. The strategy $|0\rangle$
consists in leaving the Trent's qubits intact. These moves form
the controlled--swap gate \cite{NC}. Her opponent Bob wins only if
after the game Trent learns that his qubit
$|\psi_{T}\hspace{-.1em}\rangle$ has not been changed.

\begin{figure}[h]
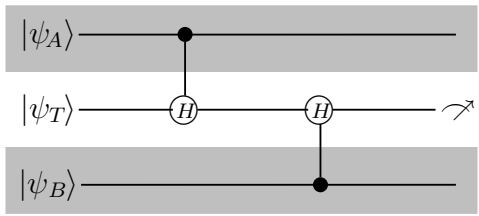

\begin{center}
\phantom{a}\vspace{11ex} \psset{linewidth=.7pt}
\pscustom[linecolor=white,fillstyle=solid,fillcolor=lightgray]{%
\psline(-3.3,1.5)(-3.3,1.5) \psline(-3.3,2.4)(3,2.4)
\psline(3,2.4)(3,1.5) }
\rput(-2.7,2){\rnode{A}{$|\psi_{\negthinspace
A}\rangle$\hspace{1pt}}} \cnode*(-.9,2){.1}{B}
\rput(2.7,2){\rnode{C}{}}
\rput(-2.7,1){\rnode{D}{$|\psi_{T}\hspace{-.1em}\rangle$\hspace{1pt}}}
\rput(-.9,1){\rnode{E}{\circh}} \rput(.9,1){\rnode{F}{\circh}}
\rput(2.7,1){\rnode{G}{\hspace{.15em}\meter}}
\pscustom[linecolor=white,fillstyle=solid,fillcolor=lightgray]{%
\psline(-3.3,.5)(-3.3,.5) \psline(-3.3,-.4)(3,-.4)
\psline(3,-.4)(3,.5) }
\rput(-2.7,0){\rnode{H}{$|\psi_{\hspace{.15em}\negthinspace
B}\hspace{-.05em}\rangle$\hspace{1pt}}} \cnode*(.9,0){.1}{I}
\rput(2.7,0){\rnode{J}{}} \ncline[nodesep=0pt]{-}{A}{C}
\ncline[nodesep=0pt]{-}{D}{E} \ncline[nodesep=0pt]{-}{E}{F}
\ncline[nodesep=0pt]{-}{F}{G} \ncline[nodesep=0pt]{-}{H}{J}
\ncline[nodesep=0pt]{-}{B}{E} \ncline[nodesep=0pt]{-}{F}{I}
\end{center}
\caption{Quantum identification game constructed from two
controlled--Hadamard gates (Wiesner's banknote).} \label{goresyt2}
\end{figure}

To win Bob must always begin with with a strategy identical to the
one used by Alice. If there is no coordination of moves between
Alice and Bob the probability of Bob's success exponentially
decreases with growing number of sub--games being played and is
negligible even for a small number of sub--games. Although Alice
and Bob's strategies are classical eavesdropping is not possible
if Trent uses arbitrary polarizations
$|\psi_{T}\hspace{-.1em}\rangle= |0\rangle+z\,|\text{I}\rangle$,
$z\in\overline{\mathcal{C}}\simeq S_2$ (in the projective
nonhomogeneous coordinates). This game can be quantized by
elimination of the ancillary qubit
$|\psi_{T'}\negthinspace\rangle$. Then Alice and Bob's strategies
should be equivalent to controlled-Hadamard gates. (The reader can
easily represent the controlled-Hadamard gates in terms of the
alliance $CNOT$ and 1-qubit tactics, cf. \cite{NC}.) In this case
Trent's qubit is changed only if Alice adopts the strategy
$|\text{I}\rangle$ that result in $|\psi_{T}\hspace{-.1em}\rangle=
|0\rangle+z\,|\text{I}\rangle\longrightarrow
|0\rangle+\frac{1-z}{1+z}\,|\text{I}\rangle$  (quantum Fourier
transform), see Fig\mbox{.} \ref{goresyt2}. The original Wiesner's
idea was to encode the secret values of
$|\psi_{T}\hspace{-.1em}\rangle$ that result from Alice moves in
the series of sub--games on an otherwise numbered banknote. In
addition, the issuer  Trent takes over the role of Alice and
records the values of $|\psi_{T}\hspace{-.1em}\rangle$ and
$|\psi_{\negthinspace A}\rangle$ with the label being the number
of the banknote.  The authentication of the banknote is equivalent
to a success in the game when Bob's strategy is used against that
recorded by Trent (if Bob wins then his forgery is successful).

  The introduction of classically
impossible strategies results in better security against quantum
attack (pretending to be Alice). Eavesdropping of the state
$|\psi_{T}\hspace{-.1em}\rangle$ modified by Alice's strategy is
ineffective  even if Trend limits himself to polarizations from
the set $\{|0\rangle,|\text{I}\rangle\}$. It is possible that an
analogous reduction of qubits allows to  exponentially reduce the
complexity of quantum algorithms. Therefore quantum games may
sometimes be the only feasible alternatives if the classical
problems are computationally too complex to be ever implemented.

\section{Kernels and shells of quantum computers: quantum game model of mind}
In the former section we have put great  emphasis on distinction
between measuring qubits and qubits being measured. The later were
shaded-in in figures. Analogously to the terminology used in
computer science, we can distinguish the shell (the measuring
part) and the kernel (the part being measured) in a quantum game
that is perceived as an algorithm  implemented by a specific
quantum process. Note that this distinction was introduced on the
basis of abstract properties of the game quantum algorithm,
quantum software) and not properties of the specific physical
implementation. Quantum hardware would certainly require a lot of
additional measurements that are nor specific to the game (or
software), cf. the process of starting a one-way quantum
computer. Adherents of artificial intelligence (AI) should welcome
the great number of new possibilities offered by quantum approach
to AI (QAI). For example, consider a Quantum Game Model of Mind
(QGMM) exploring the confrontation of quantum dichotomy between
kernel and shell with the principal assumption of psychoanalysis
of dichotomy between consciousness and unconsciousness
\cite{freud}. The relation is as follows.
\begin{itemize}
\item Kernel represents the Ego, that is the conscious or more
precisely, that level of the psyche the it is aware of its
existence (it is measured by the Id).  This level is  measured due
to its coupling to the Id via the actual or latent (not yet
measured) carriers of consciousness (in our case qubits
representing strategies) \item Shell represents the Id that is not
self-conscious. Its task is monitoring (that is measuring) the
kernel. Memes, the AI viruses \cite{memy}, can be nesting in that
part of the psyche.
\end{itemize}
Memes being qutrojans, that is quantum parasitic gates  (not
qubits!) can replicate themselves (qubits cannot -- no-cloning
theorem). Very little is known about the possible threat posed by
qutrojans to the future of quantum networks. In quantum
cryptography teleportation of qubits will be helpful in overcoming
potential threats posed by qutrojans therefore we should only
worry about attacks by conventional trojans \cite{lo99}. If the
qutrojan is able replicate itself is certainly deserves the name
quvirus.  A consistent quantum mechanism of such replication is
especially welcome if quantum computers and cryptography are to
become a successful technology. In the QGMM approach external
measuring apparatus and ``bombs'' reducing (projecting) quantum
states of the game play the role of the nervous system providing
the ``organism'' with contact with the environment that sets the
rules of the game defined in terms of supplies and admissible
methods of using of tactics and pay-offs \cite{komp}. Contrary to
the  quantum automaton put forward by Albert \cite{albert} in QGMM
model there is no self-consciousness -- only the Ego is conscious
(partially) via alliances with the Id and is infallible only if
the Id is not infected with memes. Alliances between the kernel
and the Id (shell) form states of consciousness of QAI and can be
neutralized (suppressed) in a way analogous to the quantum
solution to the Newcomb's paradox \cite{PSN}. In the context of
unique properties of quantum algorithms and their potential
applications the problem of deciding which model of AI (if any)
faithfully describes human mind is fascinating but a secondary
one. The discussed above variant of the Elitzur-Vaidman breaker
suggests that the addition of the third qubit to the kernel could
be useful in modelling the process of forming the psyche by
successive decoupling qubits from the direct measurement domain
(and thus becoming independent of the shell functions). For
example dreams and hypnosis could take place in  shell domains
that are temporary coupled to the kernel in this way. The example
discussed in the previous section illustrates what QAI intuition
resulting in a classically unconveyable  belief might be like.
What important is,  QAI reveals more subtle properties than its
classical counterparts because it can deal with counterfactual
situations \cite{mitchison, vai} and in that sense analyze
hypothetical situations (imagination). Therefore QAI is
anti-Jourdainian: Molier's Jourdain speaks in prose without
knowing it; QAI might be unable to speak but know it would have
spoken in prose  were it possible.
\section{Conclusion}
Since the publication of G\"{o}del theorems \cite{nagel} the
opinion that human  mind dominates any conceivable computer
prevails. But in the light of quantum information processing
\cite{NC} and scepticism concerning the role of quantum phenomena
in brain processes \cite{penrose2} we  might be doomed to dreary
future of coherent states of quantum matter dominating human mind.
A new fascinating field of research has been started.\\

 {\bf Acknowledgements} This paper has been supported
by the {\bf Polish Ministry of Scientific Research and Information
Technology} under the (solicited) grant No {\bf
PBZ-MIN-008/P03/2003}.

\end{document}